\documentclass[aas2pp4,natbib209]{emulateapj}
\newcommand{\simless}
     {\ensuremath{\lower
3pt\hbox{$\rlap{\raise5pt\hbox{$\char'074$}}\mathchar"7218$}}}
\newcommand{\simgreat}
     {\ensuremath{\lower
3pt\hbox{$\rlap{\raise5pt\hbox{$\char'076$}}\mathchar"7218$}}}
\newcommand{\simgt}{\lower.5ex\hbox{$\; \buildrel > \over \sim \;$}}
\newcommand{\simlt}{\lower.5ex\hbox{$\; \buildrel < \over \sim \;$}}

\shorttitle{Dust continuum and polarization in W51 North}
\shortauthors{Tang et al.}

\begin{document}
\title{Dust continuum and Polarization from Envelope to Cores in Star Formation: A Case Study in the W51 North region}
\author{Ya-Wen Tang\altaffilmark{1,2,3}, Paul T. P. Ho\altaffilmark{3,4}, Patrick M. Koch\altaffilmark{3}, Stephane Guilloteau\altaffilmark{1,2}, and Anne Dutrey\altaffilmark{1,2}}

\altaffiltext{1}{Universit\'{e} de Bordeaux, Observatoire Aquitain des Sciences de l'Univers, 2 rue de l'Observatoire, BP 89, F-33271 Floirac Cedex, France}
\altaffiltext{2}{CNRS, UMR 5804, Laboratoire d'Astrophysique de Bordeaux,
2 rue de l'Observatoire, BP 89, F-33271 Floirac Cedex, France}
\altaffiltext{3}{Academia Sinica, Institute of Astronomy and Astrophysics, Taipei, Taiwan}
\altaffiltext{4}{Center for Astrophysics, 60 Garden Street, Cambridge, MA 02138}

\email{ywtang@asiaa.sinica.edu.tw}

\begin{abstract}

We present the first high-angular resolution (up to 0$\farcs$7, $\sim$5000 AU) polarization and thermal dust continuum images toward the massive star-forming region W51 North.
The observations were carried out with the Submillimeter Array (SMA) in both the subcompact (SMA-SubC) and extended (SMA-Ext) configurations at a wavelength of 870 $\mu$m.  
W51 North is resolved into four cores (SMA1 to SMA4) in the 870 $\mu$m continuum image. 
The associated dust polarization exhibits more complex structures than seen at lower angular resolution.  
We analyze the inferred morphologies of the plane-of-sky magnetic field (B$_{\bot}$) in the SMA1 to SMA4 cores and in the envelope using the SMA-Ext and SMA-SubC data.
These results are compared with the B$_{\bot}$ archive images obtained from the CSO and JCMT. 
The polarization percentage is about 1 \% to 4 \%, and it is found to decrease with higher intensity in our SMA images, which is a similar trend as previously reported in the CSO and JCMT data. A correlation between dust intensity gradient position angles 
($\phi_{\rm \nabla I}$) and magnetic field position angles ($\phi_{\rm B}$) is found in the CSO, JCMT and both SMA data sets. 
This correlation is further analyzed quantitatively. 
A systematically tighter correlation between $\phi_{\rm \nabla I}$ and $\phi_{\rm B}$ is found in the cores, whereas the correlation decreases in outside-core regions. 
Magnetic field-to-gravity force ratio ($\Sigma_{\rm B}$) maps are derived using the newly developed polarization - intensity
gradient method by \citet{Koch+etal_2012a}. 
We find that the force ratios tend to be small ($\Sigma_{\rm B}\simlt 0.5$) in the cores in all 4 data sets. In regions outside of the cores, the ratios increase or the field
is even dominating gravity ($\Sigma_{\rm B} > 1$).
This possibly provides a physical explanation of the tightening correlation between
$\phi_{\nabla \rm I}$ and $\phi_{\rm B}$ in the cores: the more the B field lines are dragged and aligned by gravity, the tighter the correlation is. 
Finally, we propose a schematic scenario for the magnetic field in W51 North to interpret the four polarization observations at different physical scales.

\end{abstract}


%
\section{Introduction}\label{sec:introduction}

Stars form in giant molecular clouds under the threads of turbulence \citep[see review by ][]{MacLow+etal_2004} and large-scale magnetic (B) fields \citep[e.g.][]{Crutcher_1999}.  
Theoretically, the significance of the B field influences how structures are formed, such as the density contrast within structures \citep{Kowal+etal_2007}, the star formation rate \citep{Vazquez+etal_2005,Price+etal_2008} and the suppressed fragmentation \citep{Price+etal_2007,Hennebelle+etal_2011}.
However, the B fields in star-forming clouds are not well constrained observationally, because they are difficult to detect.
The Zeeman effect on spectral lines can trace the B field strength along the line of sight \citep[e.g.][]{Fish2006,Vlemmings+etal_2006,Falgarone+etal_2008,Surcis+etal_2012}.
The B field properties on the plane of sky, B$_{\bot}$, rely on polarization studies.

Dust grains are thought to be partly aligned with their minor axes to the magnetic field lines. Several alignment mechanisms have been studied \citep[see review by][]{Lazarian2000}, and radiative torques are likely to be responsible for the dust alignment \citep{Draine1996,Draine1997,Cho2005,Lazarian&Hoang_2007}. 
Due to the difference in emissivity along the major and minor axes of the elongated dust grains, the thermal emission from dust grains can therefore be partly linearly polarized perpendicular to the field lines \citep{Hildebrand1988}. 
The B$_{\bot}$ morphology can therefore be inferred by rotating the
detected polarization of the thermal emission by 90$\degr$.  
Linear polarization of thermal dust emission has been imaged at far infrared and at submillimeter (submm) in several star-forming clouds \citep[e.g.,][]{Greaves+etal_1994,Chrysostomou+etal_2002,Vaillancourt+etal_2008,Dotson+etal_2010,Matthews+etal_2009}.
To further resolve the star-forming cores, higher angular resolution observations of the millimeter (mm) and submm continuum emission are powerful, because the continuum is mostly optically thin at these wavelengths.
Both massive and low mass star-forming cores have been imaged in linear polarization with an angular resolution of a few arcsecond in the mm and submm regime \citep[e.g.,][]{Lai+etal_2001,Girart+etal_2006,Tang+etal_2009a,Tang+etal_2009b,Rao+etal_2009,Tang+etal_2010}.
W51 North is another good candidate to study the massive star formation processes. 
It has strong dust emission at mm wavelengths.
This increases the possibilities to detect polarization, because the linear polarization percentage is typically only a few percent in star-forming regions and therefore, difficult to measure. 

The W51 North region is active in star formation. It lies within the HII region complex G49.5-0.4 in W51 A, west of the W51 giant molecular cloud. 
It is located 5$-$8 kpc away in the Sagittarius spiral arm
\citep{Genzel+etal_1981,Imai+etal_2002,Xu+etal_2009}. In this paper, we adopt a distance of 7 kpc. 
At radio wavelengths (1.3 cm and 3.6 cm), an edge-brightened cometary HII region, W51 d,
and an ultracompact HII (UCHII) region, W51 d2, \citep{Gaume+etal_1993} were detected.
The brightest mm source is 2$\arcsec$ to the east of W51 d2 \citep{Zhang+etal_1998,Sollins+etal_2004,Zapata+etal_2008}. 
Associated with this mm source, a group of H$_{2}$O, OH and SiO masers were detected, and it has been also named as "the dominant center" \citep{Schneps+etal_1981}.
In this dominant center, infall signatures have been detected \citep{Sollins+etal_2004, Zapata+etal_2008,Zapata+etal_2009}.  
The outflows from the dominant center were reported to have position angles between 105$\degr$ and 150$\degr$ \citep{Eisner+etal_2002, Zhang+etal_1998,Zapata+etal_2009}. 
An embedded protostar candidate OKYM 1 \citep[also called KJD 3,][]{Kraemer+etal_2001},  1$\arcsec$ to the north-east of the dominant center, was reported \citep{Okamoto+etal_2001, Barbosa+etal_2008}. 
This region is, thus, a very active star-forming complex where several regions
are in different evolutionary stages.

%
\begin{figure*}[!ht]
\includegraphics[scale=0.5]{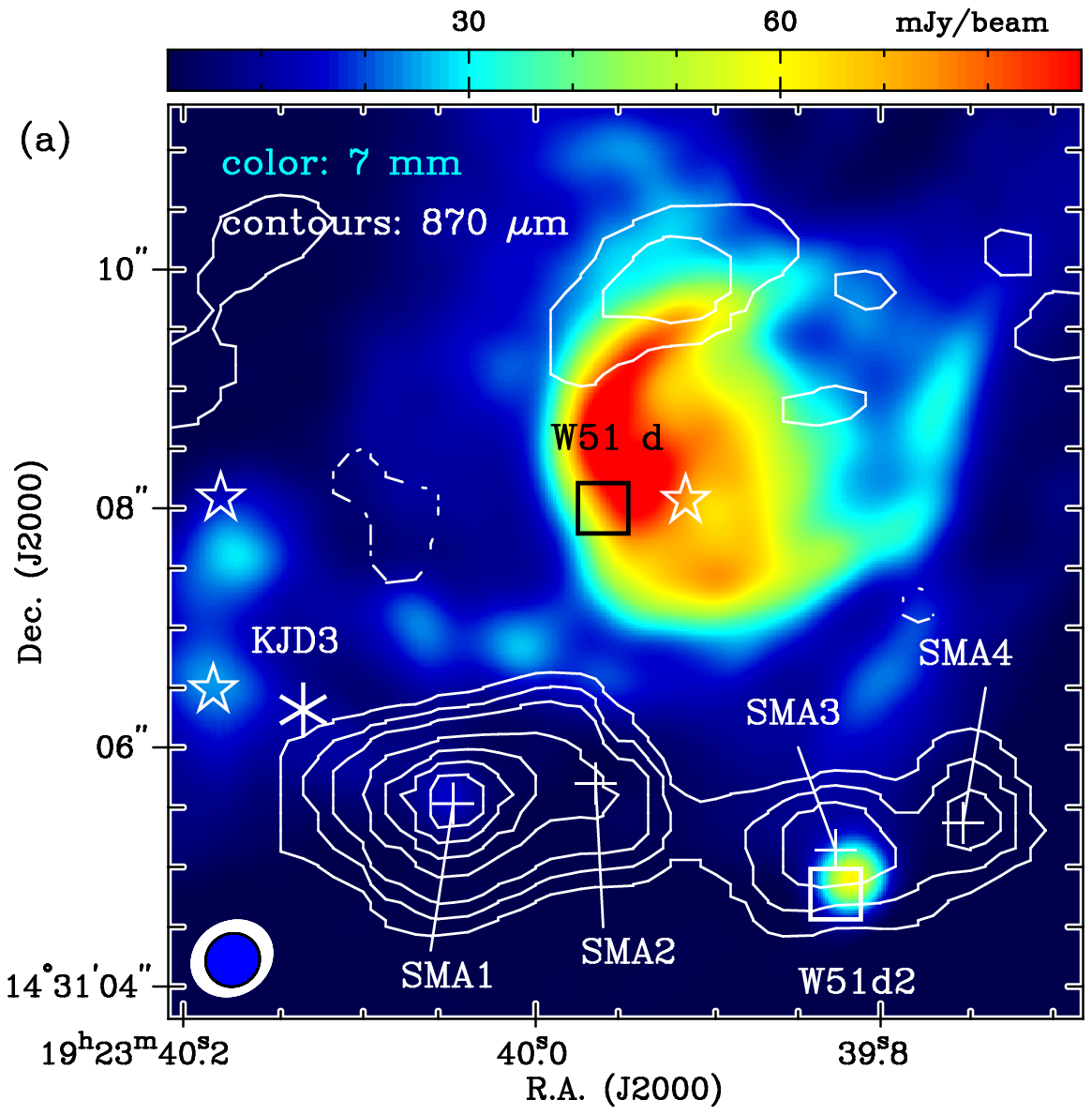}
\includegraphics[scale=0.5]{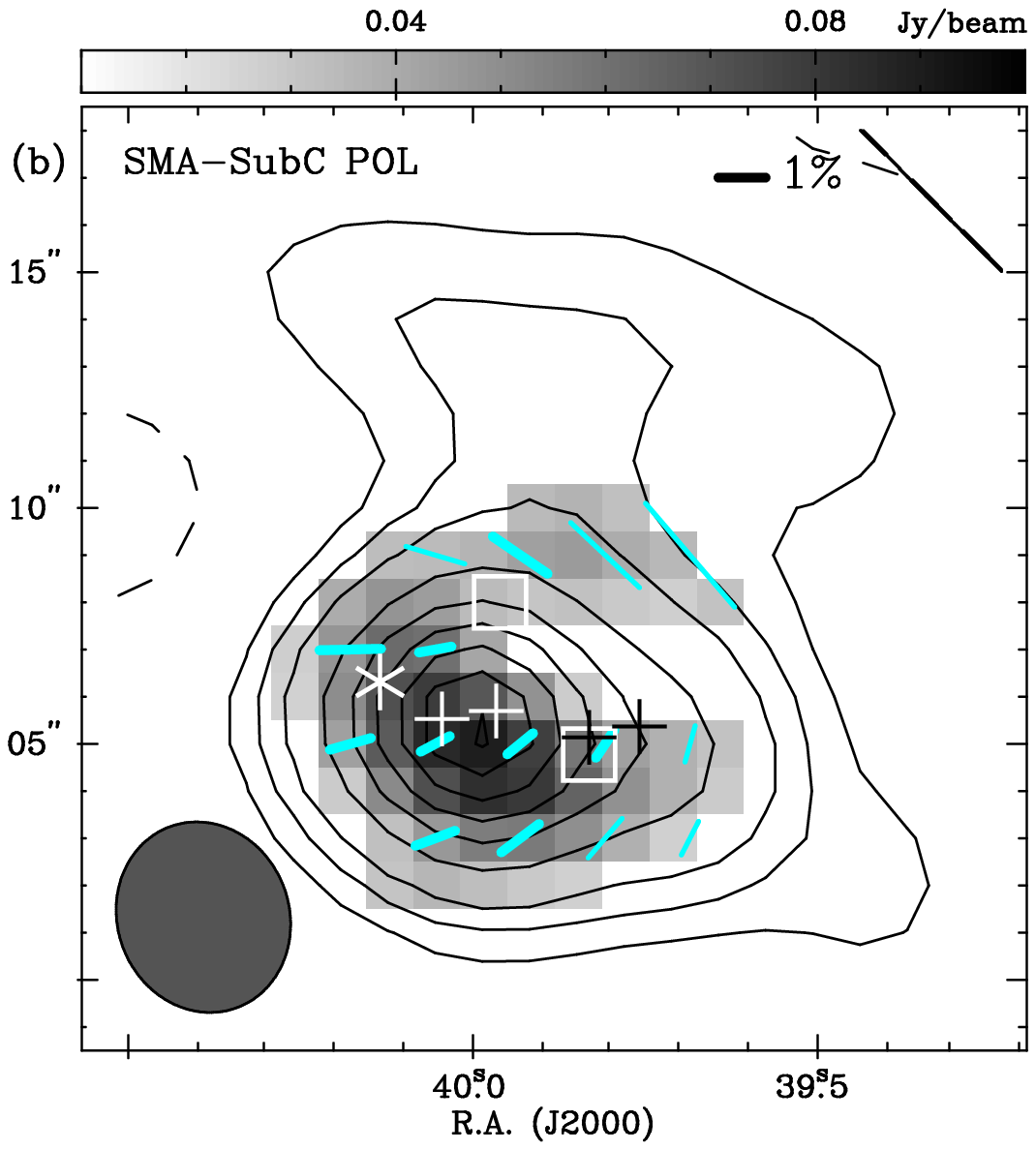}
\includegraphics[scale=0.5]{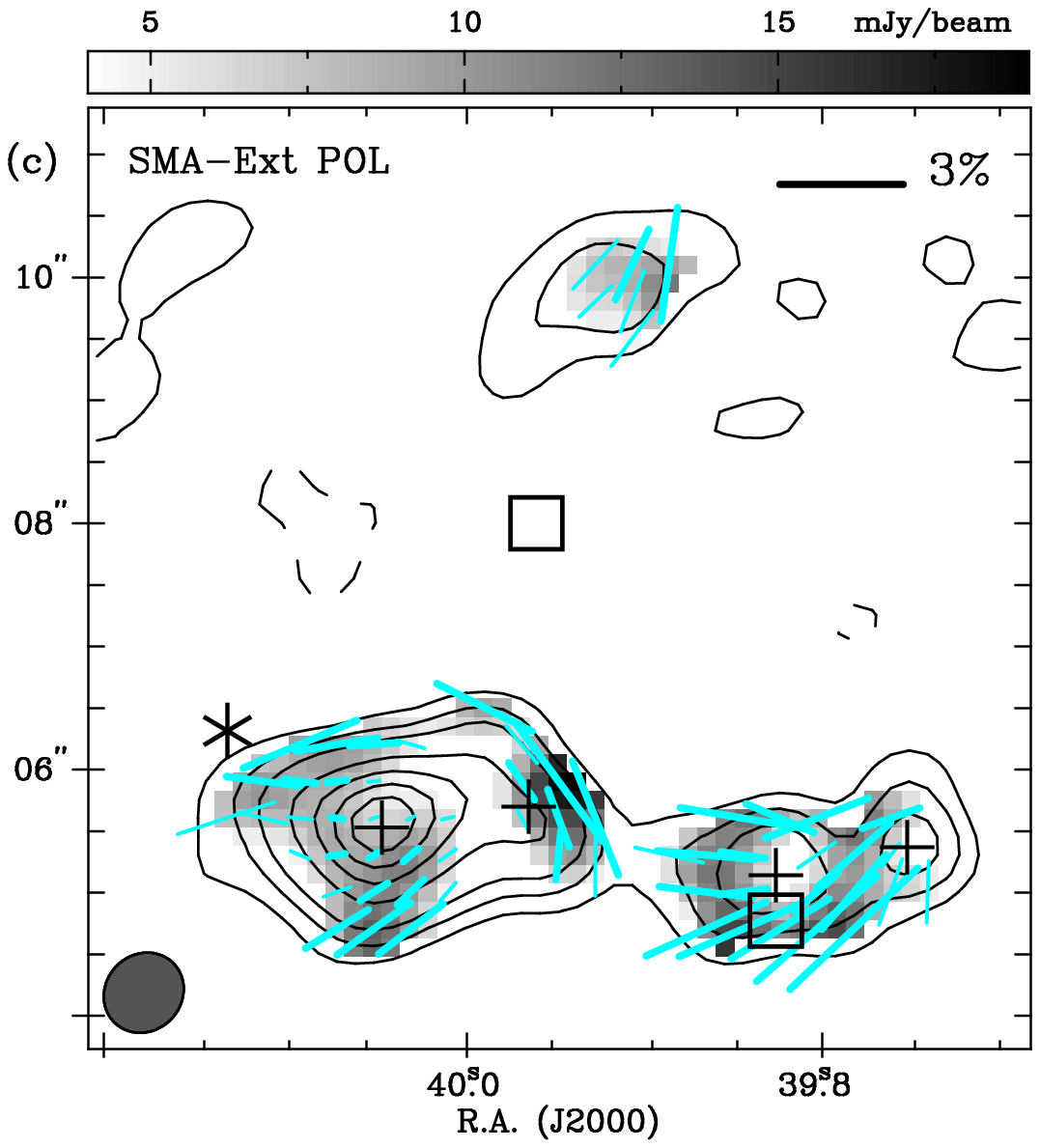}
\caption[continuum maps.]{(a): Continuum map at 870 $\mu$m obtained from SMA-Ext and continuum map at 7 mm from EVLA \citep[image adopted from][]{Zapata+etal_2008}  shown in white contours and color scale, respectively. 
The small squares mark the position of the edge-brightened
cometary/shell-like HII region W51 d and W51 d2  \citep{Gaume+etal_1993}. The infrared source KJD3 (also called OKMY 1) is marked as $\ast$. The stars mark the identified O stars by Okamoto et al. (2001). The pluses mark the resolved submillimeter sources SMA 1 to 4. (b) and (c): Polarization (segments) and polarized intensity (grey scale) from the subcompact track (panel b) and extended track (panel c). The lengths of the polarization segments are proportional to the polarization percentages.
The plotted thick and thin segments are above 3 and between 2 to 3 $\sigma_{\rm I_{\rm p}}$, respectively. The synthesized beams are shown as ellipses in the lower-left corners. Contours are at -3 (dash) and 3, 6, 10, 20, 30, 40, ... $\times$ 0.06 and 0.19 Jy beam$^{-1}$ for SMA-Ext and SMA-SubC, respectively.}
\label{fig:cont}
\end{figure*}
%
%

Linear polarization of thermal dust continuum toward W51N was detected in the submm regime with single dish telescopes. 
At wavelengths of 350 $\mu$m with an angular resolution of 20$\arcsec$,
\citet{Dotson+etal_2010} show that the polarization is mainly uniform across W51 North.
In contrast, at 850 $\mu$m, both \citet{Chrysostomou+etal_2002} and \citet{Matthews+etal_2009} show that there are more variations in the polarization orientations. 
Since there is probably multiple ongoing star formation in W51 North, 
higher angular resolution measurements of the polarization are important to reveal the physical conditions at this scale.
Hereafter, the data obtained from \citet{Dotson+etal_2010} and \citet{Chrysostomou+etal_2002} are called CSO and JCMT, respectively.  

Here, we report the new Submillimeter Array (SMA) 870 $\mu$m continuum images and the detection of linearly polarized emission of dust continuum at an angular resolution up to 0$\farcs$7 toward the W51 North.
Polarization percentages and position angles, and the correlation between position angle and intensity are studied. 
With the goal of providing a more complete scenario of the role of the magnetic field over several scales, we further compare our new polarization results at a scale of 60 mpc with the published polarization images at a scale of a few pc at the submm regime from CSO and JCMT.
The observations and data reduction are described in section \ref{sec:observation}.
The observational results are shown in section \ref{sec:result}.
In section \ref{sec:discussion}, we will present the detailed quantitative analyses of the polarization of our new observational results, and we will provide a physical explanation of the derived values. We then propose a schematic scenario for the magnetic field in W51 North.
We draw conclusions and summarize in section \ref{sec:summary}.

%
%
\section{Observations and Data Reduction}\label{sec:observation}
The observations were carried out using the SMA \citep{Ho+etal_2004} \footnote{The Submillimeter Array is a
joint project between the Smithsonian Astrophysical Observatory
and the Academia Sinica Institute of Astronomy and Astrophysics
and is funded by the Smithsonian Institution and the Academia
Sinica.} in the extended and subcompact configuration, respectively. 
In the following, the results obtained from the extended and the subcompact configuration are called SMA-Ext and SMA-SubC, respectively.
The synthesized beams (angular resolutions), $\theta_{\rm syn}$, of SMA-Ext and SMA-SubC are $\sim$ 0$\farcs$7 and 4$\arcsec$, respectively.
In both observations, the local oscillator frequency was tuned to 341.482 GHz (870 $\mu$m).
The observational details are listed in Table 1.

%
\begin{deluxetable}{l | l l}[!h]
\tablecaption{Observational parameters} \tablewidth{0pt}
\tablehead{ \colhead{Parameter} & \colhead{SMA-Ext} &
\colhead{SMA-SubC}}
\startdata

Date & 2008 Jul 13 & 2009 Sep 11 \\
Available Antennae & 7 & 6 \\
Gain Calibrator & 1751+096, 1925+211 &  1751+096 \\
Flux Calibrator & Titan & Uranus \\
Bandpass/Pol Calib. & 3c454.3 & 3c454.3 \\
Baseline range ($k\lambda$) & 30 to 262 & 10 to 30\\
$\sigma_{\rm I}$ (Jy beam$^{-1}$) & 0.06 & 0.19 \\
$\sigma_{\rm I_{\rm p}}$ (mJy beam$^{-1}$) & 3  & 15 \\
$\theta_{\rm syn}$ & 0$\farcs$7$\times$0$\farcs$6 &  4$\farcs$1$\times$3$\farcs$6 \\
P.A. of $\theta_{\rm syn}$ & -57$\degr$ & 20$\degr$ 
\enddata
\tablecomments{$\sigma_{\rm I}$ and $\sigma_{\rm I_{p}}$ are the noise levels of Stokes $I$ and the polarized intensity ($I_{\rm p}$), respectively.}
\end{deluxetable}
%
%

%
\begin{deluxetable*}{l | l l l l l  l l}[!h]
\tablecaption{Dust components \label{tab:dust}} \tablewidth{0pt}
\tablehead{ \colhead{Name} & \colhead{R.A.} &
\colhead{Dec.} & \colhead{Flux (Jy)} & \colhead{Deconvolved Size ($\arcsec\times\arcsec$), PA(\degr)} & \colhead{$\langle\phi_{\rm pol}\rangle$} & \colhead{$\sigma_{\phi_{\rm pol}}$} & \colhead{Note}}
\startdata
SMA1 & 19:23:40.047 & 14:31:5.53 & 8.69$\pm$0.27 & (0.98$\pm$0.03)$\times$(0.68$\pm$0.02), -89$\pm$3 & -25 & 51 & dominant center\\
SMA2 & 19:23:39.970  & 14:31:5.70  & 2.59$\pm$0.24 & (1.12$\pm$0.07)$\times$(0.28$\pm$0.11), 39$\pm$3 & 4 & 30 & -\\
SMA3 & 19:23:39.828 & 14:31:5.14 & 2.41$\pm$0.27 & (1.07$\pm$0.09)$\times$(0.59$\pm$0.09), -88$\pm$7 & -11 & 61 & W51d2 \\
SMA4 & 19:23:39.756 & 14:31:5.37 & 1.40$\pm$0.22 & (0.79$\pm$0.12)$\times$(0.49$\pm$0.13), -23$\pm$21& -30 & 24 & -
\enddata
\tablecomments{Resolved components of the east-west dust ridge in the SMA-Ext data. $\langle\phi_{\rm pol}\rangle$ and $\sigma_{\phi_{\rm pol}}$ are the mean and the standard deviation of the polarization position angles ($\phi_{\rm pol}$), respectively.
Deconvolved sizes for SMA1 to SMA4 are obtained by fitting 2D-Gaussians. }
\end{deluxetable*}
%
%

The SMA-Ext observation was carried out
along with W51 e2/e8. These two sources shared the same
calibrators and were reduced and imaged in the same way. 
The details of the observation and data reduction are described in Tang et
al. (2009). In brief, the SMA-Ext data were reduced and imaged in MIRIAD following standard procedures. 
An additional calibration on the instrumental polarization (also called the leakage terms) was derived using the calibrator 3c454.3. 
The largest size scale which could be sampled in the extended track was $\sim$8$\arcsec$ (0.3 pc). For the subcompact track, the data were reduced with MIR and 
then imaged using MIRIAD. 

The position angle of the polarization, $\phi_{\rm pol}$, is calculated from: $\phi_{\rm pol}$ = $\frac{1}{2}$ tan$^{-1} \frac
{U}{Q}$, increasing counter-clockwise in the range $-$90$
\degr$ to 90$\degr$. 
The strength, $I_{\rm p}$, and percentage, $P(\%)$,
of the linearly polarized emission are calculated from: $I_{\rm
p}^{2}$ = $Q^{2}$ + $U^{2}$ - $\sigma_{\rm Q,U}^{2}$ and $P(\%)$ = $I_{\rm
p}/I$, respectively. Here, $\sigma_{\rm Q,U}$ is the noise level of
the Stokes $Q$ and $U$ images, and it is also the bias correction due
to the positive measure of $I_{\rm p}$ (Leahy 1989; Wardle \&
Kronberg 1974). 

The primary beam (field of view) of the SMA at 345 GHz is $\sim$30$\arcsec$. 
All of the presented SMA maps in the following have been corrected for the primary beam attenuation. The presented maps were constructed with natural 
weighting. The phase center is at Right Ascension (J2000)=19$^{h}$23$^{m}$40$.05^{s}$, Declination (J2000)=14$\degr$31$\arcmin$5$\farcs$0.

%
%
\section{Results}\label{sec:result}
%
%
\subsection{Continuum Emission}
\subsubsection{Dense Structures: SMA 1 to SMA4}
The 870 $\mu$m continuum emission is resolved in both SMA-SubC and SMA-Ext. With 
$\theta_{\rm syn}$ of 4$\arcsec$ (SMA-SubC), 
the 870 $\mu$m continuum emission is elongated in the east-west (EW) direction with an extension toward the north (Figure \ref{fig:cont}b). 
This is consistent with the 2 mm continuum map \citep{Zhang+etal_1998} and the thermal NH$_{3}$ emission \citep{Zhang&Ho_1995}. 
With $\theta_{\rm syn}$ of 0$\farcs$7 (SMA-Ext), the EW ridge is further resolved into SMA1 to SMA4, as labeled in Figure \ref{fig:cont}a. 
The coordinates and flux densities of SMA1 to SMA4 are determined by fitting 2-D Gaussians. They are listed in Table \ref{tab:dust}.  

The brightest 870 $\mu$m continuum peak is SMA1, which is consistent with the location of the brightest peak at 7 mm \citep{Zapata+etal_2009} and at 2 mm \citep{Zhang+etal_1998}.
SMA2, SMA3 and SMA4 are resolved for the first time in the mm/submm regime. 
For SMA2, although there is no previous report of detection, 
it can be seen at 1.3 mm \citep[Figure 1 in][]{Zapata+etal_2008}. 
Associated with SMA3 is the UCHII region W51 d2, 0$\farcs$5 to the south, and a NH$_{3}$ maser \citep{Mauersberger+etal_1987,Wilson+etal_1991,Gaume+etal_1993}. 
Among these 4 resolved SMA sources, SMA3 is the most evolved source due to its free-free emission, which suggests the existence of ionized gas. 
The existence of masers in SMA1 and SMA3 suggests that these two locations are active in star formation.

\subsubsection{Flux Density and Mass}
The flux density detected with SMA-SubC is 27$\pm$3 Jy. 
Comparing to the single dish measurement at the same location of 44 Jy (Chrysostomou et al. 2002), about 60\% of the flux density is recovered in SMA-SubC. 
The compact structures (SMA1 to SMA4) identified from SMA-Ext have a total flux density of 15$\pm$1 Jy over the EW dust ridge. 
Comparing to the total emission of 44 Jy, these compact structures make for about 34\% of the total emission. 
The majority of the emission is, thus,  in the extended structure which is filtered out by the sampling of longer baselines.

The free-free continuum emission is 2.5 mJy at 3.6 cm toward W51 d2 \citep[named as SMA3 in this paper;][]{Gaume+etal_1993}, and it will be 1.7 mJy or 16 mJy when extrapolated to 870 $\mu$m, assuming an optically thin spectral index ($\alpha$) of $-0.1$ or an optically thick $\alpha$ of 0.6, respectively. 
While the flux density of SMA3 is 2.4 Jy at 870$\mu$m, the contribution from free-free emission is negligible. The 870$\mu$m continuum emission along SMA1 to SMA4 is, thus, mainly from thermal dust emission.
For the extended HII region W51 d, the flux density is 1 Jy at 1.3 cm, and it will be 0.8 Jy or 5 Jy at 870 $\mu$m assuming $\alpha$ of $-$0.1 or 0.6, respectively. At this location, the detected intensity at 870 $\mu$m with SMA-SubC is $\sim$5 Jy beam$^{-1}$.
Thus, the traced extension to the north at 870 $\mu$m could have a significant contribution from the free-free emission, if this free-free emission is not optically thin between  the wavelength of 1.3 cm and 870 $\mu$m.

The total gas mass, $M_{\rm gas}$, can be estimated from the dust continuum, assuming a gas temperature of 100 K \citep{Zhang+etal_1998}, a dust opacity $\kappa_{\lambda=870\mu m}\propto \lambda^{-2}\approx$ 0.4 cm$^{2}$ g$^{-1}$ \citep[following the standard equation in][]{Agladze+etal_1996}, and a gas-to-dust ratio of 100.
For SMA1, $M_{\rm gas}$ is $\sim$130 $M_{\sun}$, roughly consistent with the value reported in \citet{Zhang+etal_1998} of 100 M$_{\sun}$ and \citet{Zapata+etal_2008} of 90 M$_{\sun}$. For SMA 2, SMA 3 and SMA 4, $M_{\rm gas}$ is 39, 36 and 21 M$_{\sun}$, respectively.
We note that the estimated mass can vary by a factor of a few due to the uncertainty in $\kappa_{\lambda}$, the assumed dust temperature, 
the gas-to-dust ratio and the distance to the source.

%
\begin{figure*}[!ht]
\begin{center}
\includegraphics[scale=1]{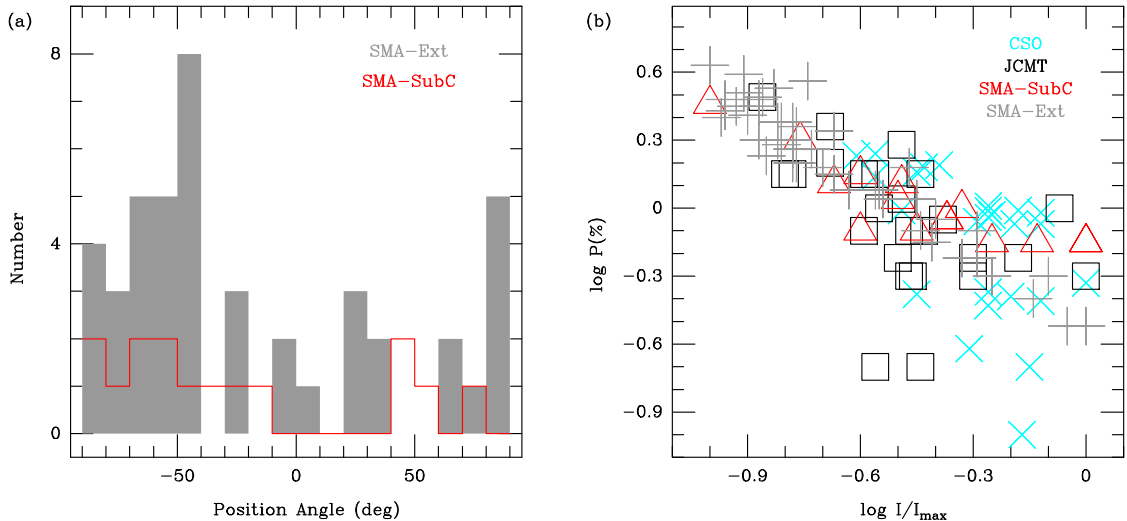}
\caption{(a) Distributions of detected polarization position angles ($\phi_{\rm pol}$). (b) Polarization percentage ($P$\%) versus normalized
Stokes $I$ dust intensity, $I/I_{max}$.}
\label{fig:histo_pa}
\end{center}
\end{figure*}
%
%

%
\begin{deluxetable}{r r c r c r c}[!h]
\tabletypesize{\footnotesize}
\tablecaption{Polarization data detected with the SMA} \tablewidth{0pt}
\tablehead{ \colhead{$\triangle$R.A.} & \colhead{$\triangle$Dec.} &
\colhead{$P$} & \colhead{$\phi_{\rm pol}$} & \colhead{$\sigma_{\phi_{\rm pol}}$}& \colhead{$I_{\rm p}$} & \colhead{$I$}\\
\colhead{($\arcsec$)} & \colhead{($\arcsec$)} & \colhead{($\%$)} &\colhead{($\degr$)} & 
\colhead{($\degr$)} & \colhead{(mJy/b)} & \colhead{(Jy/b)}}
\startdata
\multicolumn{6}{l}{Extended Track}\\
  0.3 &  -0.3 &   1.8 $\pm$   0.6 &   -58 &    10 &     9 &  0.48\\
  0.0 &  -0.3 &   2.2 $\pm$   0.5 &   -55 &     6 &    14 &  0.62\\
 -0.3 &  -0.3 &   2.0 $\pm$   0.6 &   -52 &     9 &     9 &  0.44\\
 -2.7 &  -0.3 &   3.2 $\pm$   0.8 &   -66 &     7 &    11 &  0.36\\
 -3.0 &  -0.3 &   3.4 $\pm$   0.6 &   -67 &     5 &    15 &  0.43\\
 -3.3 &  -0.3 &   2.8 $\pm$   0.7 &   -58 &     7 &    11 &  0.39\\
 -3.6 &  -0.3 &   3.9 $\pm$   0.8 &   -49 &     6 &    14 &  0.36\\
 -3.9 &  -0.3 &   4.3 $\pm$   1.0 &   -47 &     6 &    12 &  0.29\\
  0.3 &   0.0 &   0.7 $\pm$   0.2 &   -70 &    10 &     8 &  1.13\\
  0.0 &   0.0 &   0.8 $\pm$   0.2 &   -57 &     7 &    11 &  1.50\\
 -0.3 &   0.0 &   0.9 $\pm$   0.2 &   -47 &     8 &    10 &  1.17\\
 -2.7 &   0.0 &   2.3 $\pm$   0.6 &   -97 &     7 &    12 &  0.50\\
 -3.0 &   0.0 &   1.2 $\pm$   0.4 &   -83 &     8 &    10 &  0.80\\
 -3.6 &   0.0 &   1.4 $\pm$   0.4 &   -49 &     9 &     9 &  0.63\\
 -3.9 &   0.0 &   2.4 $\pm$   0.6 &   -41 &     8 &    11 &  0.45\\
 -4.2 &   0.0 &   1.7 $\pm$   0.7 &   -20 &    11 &     7 &  0.41\\
  0.3 &   0.3 &   0.4 $\pm$   0.1 &   -81 &     9 &     9 &  2.13\\
  0.0 &   0.3 &   0.3 $\pm$   0.1 &   -61 &     8 &     9 &  2.92\\
 -0.3 &   0.3 &   0.5 $\pm$   0.1 &   -49 &     7 &    11 &  2.30\\
 -1.5 &   0.3 &   1.2 $\pm$   0.3 &    -6 &     8 &    10 &  0.83\\
 -1.8 &   0.3 &   2.0 $\pm$   0.7 &    -0 &    10 &     8 &  0.39\\
 -2.7 &   0.3 &   2.4 $\pm$   0.6 &    85 &     7 &    12 &  0.49\\
 -3.0 &   0.3 &   1.1 $\pm$   0.3 &    82 &     8 &    10 &  0.85\\
 -3.9 &   0.3 &   2.4 $\pm$   0.6 &   -49 &     8 &    10 &  0.44\\
  1.2 &   0.6 &   2.5 $\pm$   0.9 &   -72 &    10 &     8 &  0.31\\
  0.9 &   0.6 &   1.2 $\pm$   0.4 &    77 &    10 &     8 &  0.69\\
  0.6 &   0.6 &   0.6 $\pm$   0.2 &    84 &    10 &     8 &  1.50\\
  0.3 &   0.6 &   0.3 $\pm$   0.1 &   -97 &     9 &     9 &  2.59\\
 -1.2 &   0.6 &   0.6 $\pm$   0.2 &    29 &    10 &     8 &  1.38\\
 -1.5 &   0.6 &   1.5 $\pm$   0.3 &    20 &     5 &    15 &  0.98\\
 -1.8 &   0.6 &   3.0 $\pm$   0.7 &    22 &     7 &    12 &  0.39\\
 -3.0 &   0.6 &   3.1 $\pm$   0.7 &    80 &     6 &    12 &  0.40\\
 -3.3 &   0.6 &   1.8 $\pm$   0.6 &    67 &     9 &     9 &  0.49\\
 -3.6 &   0.6 &   2.7 $\pm$   0.9 &   -69 &     9 &     9 &  0.32\\
 -4.2 &   0.6 &   1.5 $\pm$   0.5 &   -70 &     9 &     9 &  0.58\\
  0.9 &   0.9 &   1.8 $\pm$   0.5 &    83 &     8 &    10 &  0.55\\
  0.6 &   0.9 &   0.8 $\pm$   0.3 &   -97 &     9 &     9 &  1.05\\
  0.3 &   0.9 &   0.5 $\pm$   0.2 &   -83 &     9 &     9 &  1.63\\
 -1.2 &   0.9 &   1.1 $\pm$   0.3 &    35 &     7 &    11 &  1.03\\
 -1.5 &   0.9 &   3.6 $\pm$   0.5 &    36 &     4 &    19 &  0.53\\
  0.6 &   1.2 &   3.0 $\pm$   0.9 &   -67 &     8 &    10 &  0.32\\
  0.3 &   1.2 &   2.0 $\pm$   0.6 &   -81 &     8 &    10 &  0.49\\
  0.0 &   1.2 &   1.2 $\pm$   0.4 &    94 &    10 &     8 &  0.68\\
 -0.9 &   1.5 &   2.6 $\pm$   0.8 &    63 &     8 &     9 &  0.37\\
 -2.1 &   4.8 &   1.6 $\pm$   0.6 &   -22 &    10 &     8 &  0.49\\
 -1.8 &   5.1 &   1.6 $\pm$   0.6 &   -43 &    10 &     8 &  0.50\\
 -2.1 &   5.1 &   1.9 $\pm$   0.6 &   -25 &     9 &     9 &  0.45\\\hline \\ 
\multicolumn{6}{l}{Subcompact Track}\\
  0.0 &  -2.0 &   0.9 $\pm$   0.3 &   -69 &     9 &    47 &  5.15 \\
 -1.8 &  -2.0 &   1.0 $\pm$   0.3 &   -53 &    7 &    57 &  5.70 \\
 -3.6 &  -2.0 &   1.1 $\pm$   0.4 &   -41 &    10 &    43 &  3.83 \\
 -5.4 &  -2.0 &   0.8 $\pm$   0.5 &   -27 &    17 &    25 &  3.04 \\
  1.8 &   0.0 &   0.9 $\pm$   0.3 &   -74 &     9 &    47 &  5.14 \\
  0.0 &   0.0 &   0.7 $\pm$   0.1 &   -62 &    5 &    86 & 12.15 \\
 -1.8 &   0.0 &   0.7 $\pm$   0.1 &   -50 &     5 &    80 & 12.07 \\
 -3.6 &   0.0 &   0.7 $\pm$   0.2 &   -34 &     9 &    47 &  6.85 \\
 -5.4 &   0.0 &   0.8 $\pm$   0.3 &   -15 &    13 &    34 &  4.33 \\
  1.8 &   2.0 &   1.3 $\pm$   0.4 &   -88 &     9 &    50 &  3.93 \\
  0.0 &   2.0 &   0.7 $\pm$   0.2 &   -80 &     7 &    61 &  9.05 \\
  0.0 &   4.0 &   1.3 $\pm$   0.6 &    74 &    13 &    34 &  2.60 \\
 -1.8 &   4.0 &   1.4 $\pm$   0.5 &    56 &    10 &    43 &  3.05 \\
 -3.6 &   4.0 &   2.0 $\pm$   0.7 &    47 &    10 &    42 &  2.10 \\
 -5.4 &   4.0 &   2.9 $\pm$   1.2 &    41 &    12 &    35 &  1.21 
 \enddata
\tablecomments{Offset in Right Ascension ($\triangle$R.A.) and offset in Declination ($\triangle$Dec.) with respect to the phase center (19$^h$23$^m$40.05$^s$, 14${\degr}$31\arcmin 5$\farcs$0)$_{\rm J2000}$, polarization percentage ($P$), position angle of polarization ($\phi_{\rm pol}$), the uncertainty of the position angle of polarization ($\sigma_{\phi_{\rm pol}}$), polarization intensity ($I_{\rm p}$), and Stokes I intensity ($I$). $I_{\rm p}$ and $I$ are in units of mJy beam$^{-1}$ (mJy/b) and Jy beam$^{-1}$ (Jy/b), respectively. 
Data listed are above 2.5$\sigma_{\rm I_p}$ and above 2$\sigma_{\rm I_p}$ for SMA-Ext and SMA-SubC, respectively. \label{tab:pol_data}}
\end{deluxetable}
%
%

%
\subsection{Dust Polarization}
\subsubsection{Morphology}
The linear polarization is detected in most parts of the dust ridge (Figures \ref{fig:cont}b and \ref{fig:cont}c).
The presented polarization segments were gridded every 0$\farcs$3 for SMA-Ext and every 
2$\arcsec \times 1\farcs8$ for SMA-SubC, which is about half of the synthesized beam in order to present the
variations of the position angles (P.A.s). 
The detected polarization data above 2.5 $\sigma_{I_{\rm p}}$ for the SMA-Ext and above 2 $\sigma_{I_{\rm p}}$ for the SMA-SubC are listed in Table \ref{tab:pol_data}. 
We choose to set a 2$\sigma_{I_{\rm p}}$ threshold for the SMA-SubC in order to recover more polarization segments, where otherwise only 8 (out of 15) for SMA-SubC and 43 (out of 48) for SMA-Ext segments would remain with a 3$\sigma_{I_{\rm p}}$ limit.

In the SMA-SubC map (Figure \ref{fig:cont}b), there is a region without significant polarization about 2$\arcsec$ north of SMA3.
As zooming in with SMA-Ext, there are zones without significant polarization in between SMA1 and SMA2, between SMA2 and SMA3 and at the peak of SMA3 and SMA4. 
The areas without significant polarization near the intensity peaks and in between SMA1 and SMA2 may be due to complex underlying field geometries, as they appear in between two dramatically different PA regions. 
One example of a complex field in a region with low/no polarization detection is the W51 e2/e8 region. 
Observationally, the W51 e2 core was found to have low/no polarization in the 1.3 mm continuum with lower angular resolution (3$\arcsec$) by \citet{Lai+etal_2001}. This core has actually been detected and resolved in polarization at 870 $\mu$m at higher angular resolution (0.7$\arcsec$) by \citet{Tang+etal_2009b}. 

\subsubsection{Distribution Plots}
The distributions of the polarization position angles ($\phi_{\rm pol}$) are shown in Figure \ref{fig:histo_pa}a.
In the shorter baseline data (SMA-SubC), the $\phi_{\rm pol}$ are mainly in two ranges, namely around 50$\degr$ and around -60$\degr$. 
As we will discuss in section \ref{sec:field_structure}, this bimodal-like distribution reflects a mirror symmetry in 
the polarization structure.
The component at 50$\degr$ is consistent with 
the single dish polarization measurement at 350 $\mu$m with an angular resolution of about 20$\arcsec$ (Dotson et al. 2010). 
With smaller $\theta_{\rm syn}$ (SMA-Ext), the dust ridge and polarization are 
further resolved and exhibit more structures. 
In the majority of the detected patches, values for $\phi_{\rm pol}$ are around $-50 \degr$ with smooth variations within these areas
(Figure \ref{fig:cont}c). The distribution for SMA-Ext is, consequently, peaking around $-50 \degr$ in Figure \ref{fig:histo_pa}a. 
Near SMA2, the $\phi_{\rm pol}$ are very different from the $\phi_{\rm pol}$ at the other positions. The $\phi_{\rm pol}$ of these segments are more perpendicular to the intensity gradients at these positions (Figure \ref{fig:cont}c). 
Since there are no reported ongoing star formation activities here, the origin of this abrupt change in $\phi_{\rm pol}$ from
SMA1 to SMA2 and to SMA3 is not clear.
The component with $\phi_{\rm pol}$ of 50$\degr$ seen in the SMA-SubC map is not seen in the SMA-Ext map (Figure \ref{fig:cont}c). 
This suggests that this component originates from extended structures and is not sampled in the SMA-Ext data.

%
\begin{figure*}[!ht]
\includegraphics[scale=0.65]{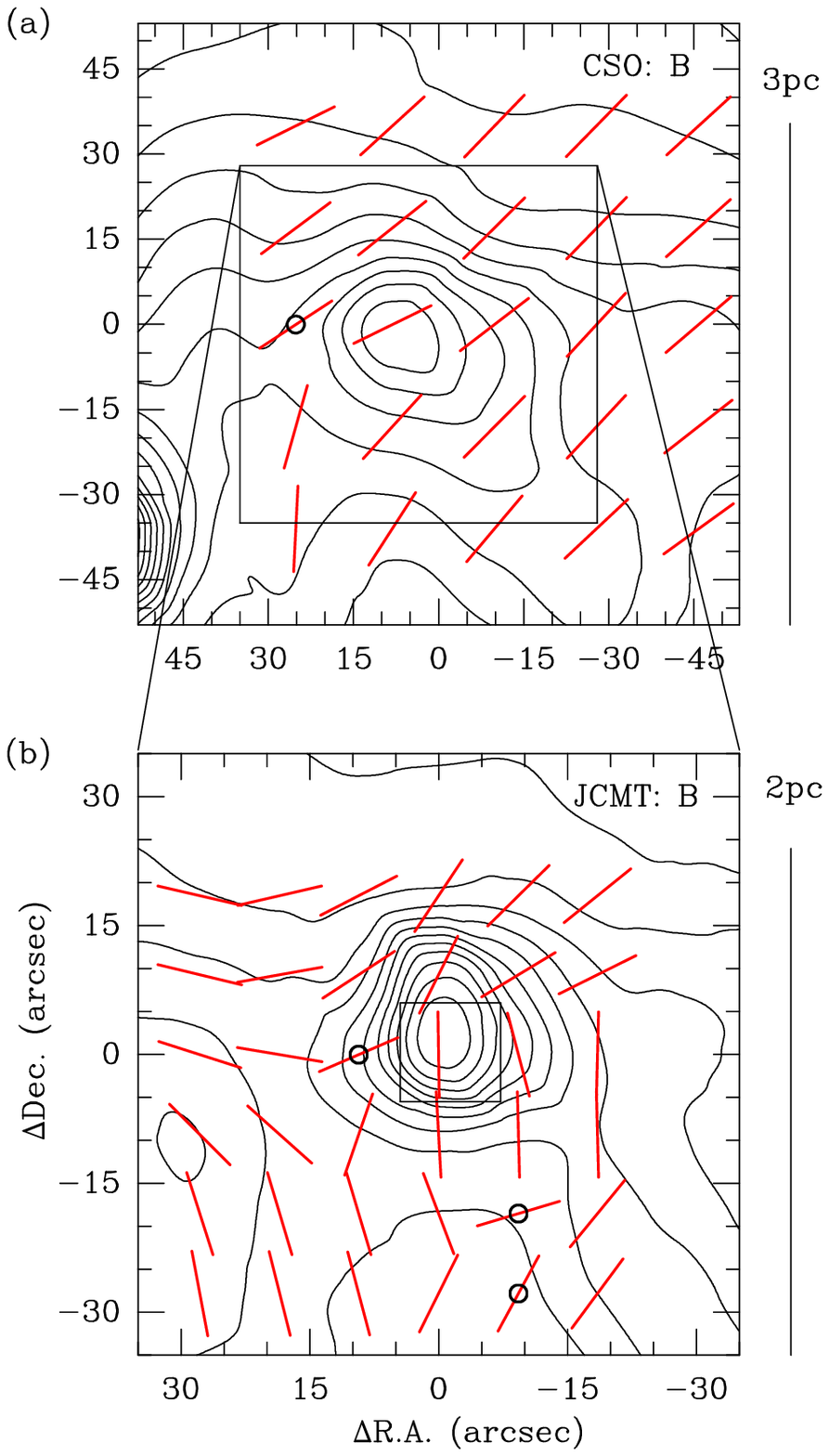}
\includegraphics[scale=1.0]{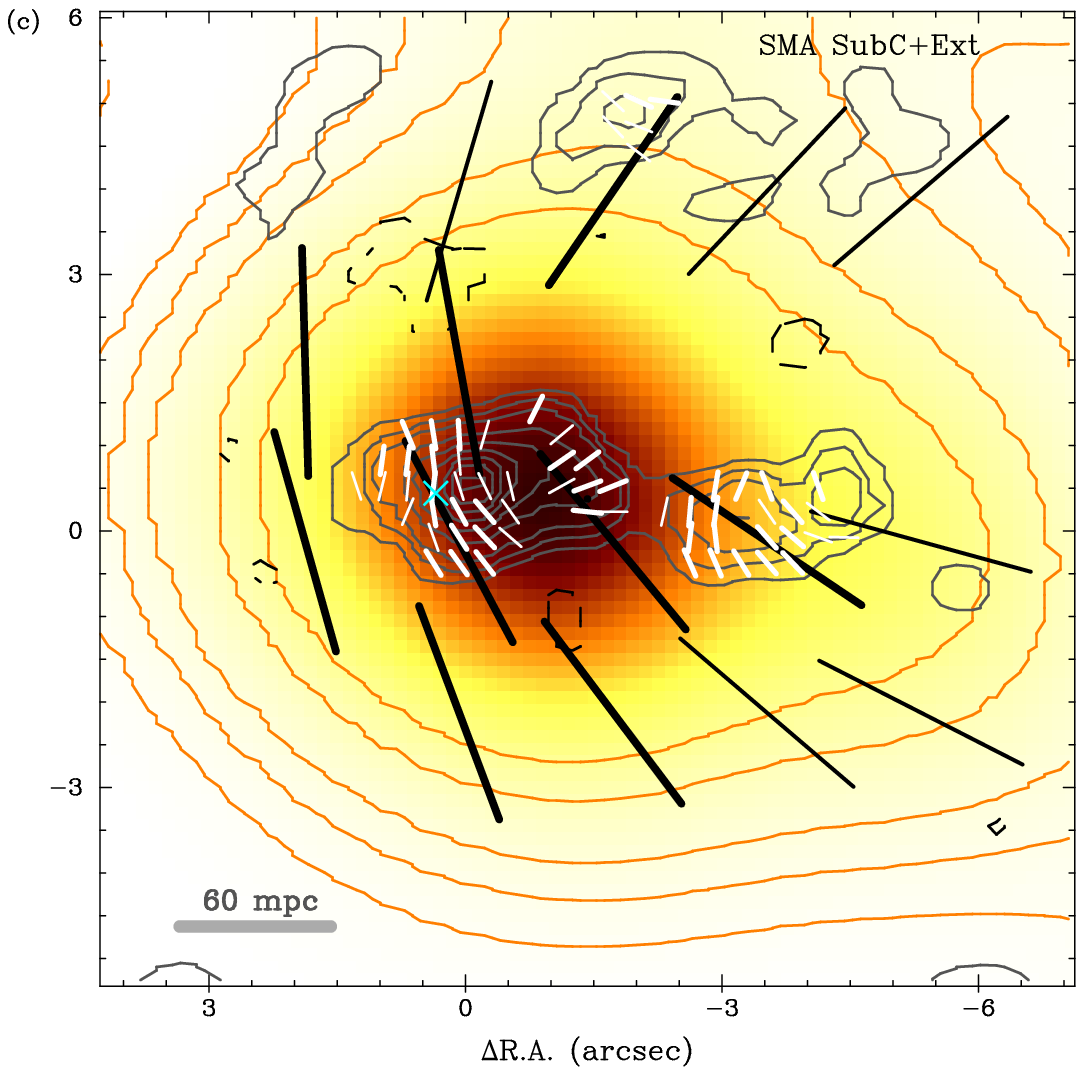}
\caption{The B field map in W51 North obtained from four different scales and/or layers (see also Figure \ref{fig:schematic}): 3 pc scale by HERTZ on the CSO in (a) and 2 pc scale by SCUBA on the JCMT in (b), and sub-pc scale by the SMA-SubC (black segments) and by SMA-Ext (white segments) in (c). In (a) and (b), the contours and the red segments label the continuum emission and the B field orientations, respectively. The small circles mark the data with larger uncertainties in position angles (15$\degr$ to 30$\degr$) as compared to the other data points.
The contours start from and step in 5$\%$ of the peak intensity detected in CSO of 694 Jy and in JCMT of 44 Jy in (a) and (b), respectively. The black square in panel (a) marks the mapping area in panel (b), and the one in panel (b) marks the mapping area in panel (c). 
In (c), the continuum emission traced with the SMA-SubC is shown in orange contours and in color scale, 
and the contours are 3, 6, 10, 20, 30 $\times$ 0.19 (Jy beam$^{-1}$).
The continuum emission detected in the SMA-Ext is plotted in grey contours in the same level as in Figure 1. 
The cyan $\times$ marks the center of the rotating toroid detected by \citet{Zapata+etal_2009}.}
\label{fig:overlay_B}
\end{figure*}
%
%

A log-log plot of the polarization percentage ($P$) versus the normalized intensity is shown in Figure \ref{fig:histo_pa}b.
$P$(\%) is in the range of $<$1 to 4.5\% and the mean $P$(\%) is 1.81\% and 1.07\% for the SMA-Ext and SMA-SubC, respectively. 
There is a general trend for both our high-angular-resolution data and the single dish data from the CSO and the JCMT: $P$ decreases with 
larger intensity.
Numerical simulations have shown that such anti-correlations can be due to a more complex $\phi_{\rm pol}$-structure in dense regions \citep{Goncalves+etal_2008}.

%
\begin{figure*}
\begin{center}
\includegraphics[scale=0.72]{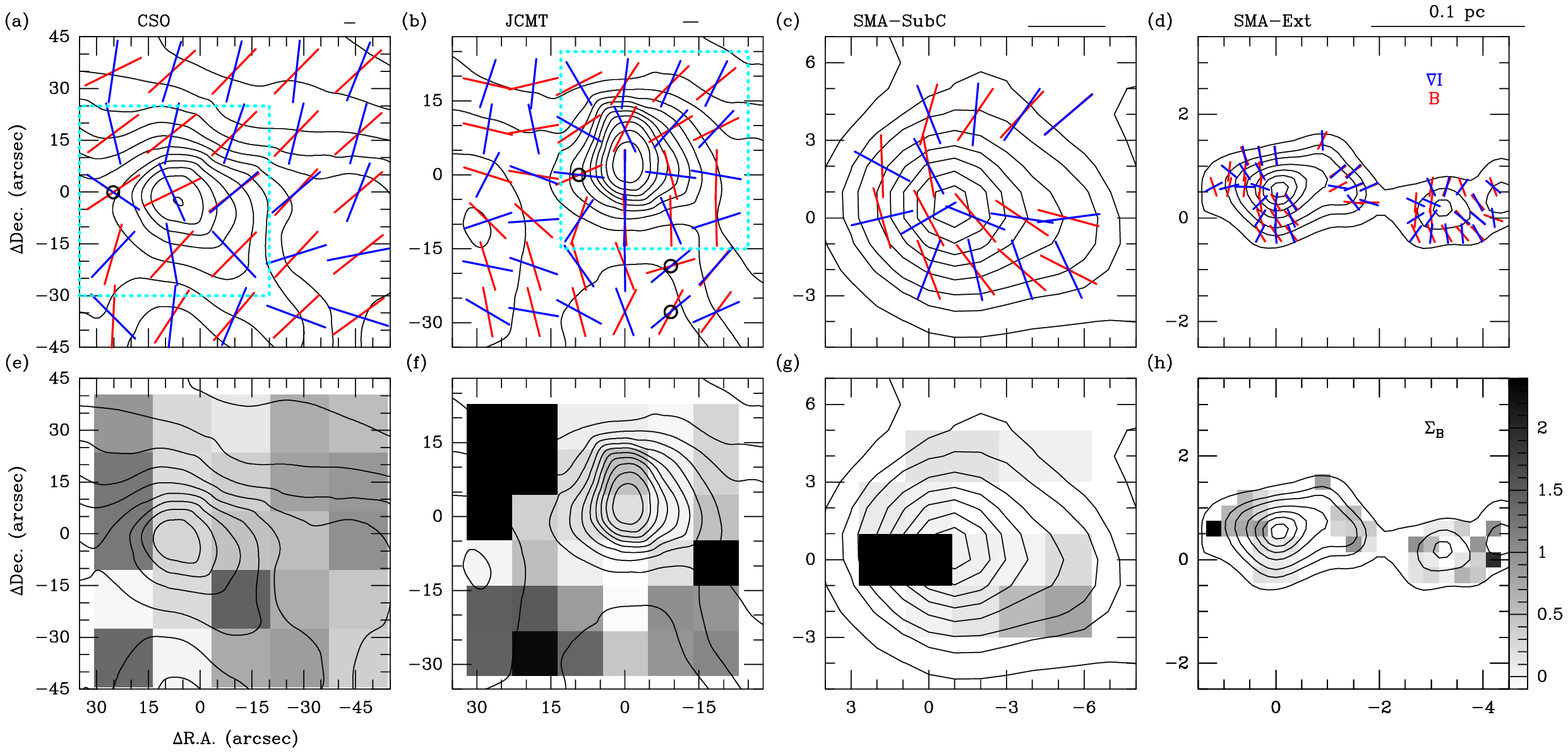}
\caption{\label{fig:analysis}
Comparison of the four different field morphologies observed with Hertz on the CSO in(a,e), SCUBA on the JCMT in (b,f), the 
SMA-SubC in (c,g) and the SMA-Ext in (d,h). In the top row (a-d), the total intensity (contours), the B field (red segments) and the intensity gradient (blue segments) orientations are shown. The derived force ratios of the B field to the gravity, $\Sigma_{\rm B}$, are shown in grey scale in the bottom row panels (e-h). Darker colors indicate that the B field is dominant ($\Sigma_{\rm B}$ $>$ 1). Lighter colors indicate that the B field plays a minor role ($\Sigma_{\rm B}$ $<$ 1). Cyan squares mark the core areas used in Table \ref{tab:analysis}. A segment representing a 0.1 pc scale-length is plotted at the upper-right corners for a reference. The small black circles in the panels 
(a) and (b) mark the data with magnetic field P.A. uncertainties between 15 to 30$\degr$.}
\end{center}
\end{figure*}
%
%

%
\section{Discussion}\label{sec:discussion}

In this section we discuss the different morphologies in the dust emission
with its associated B field from 4 different scales/layers. 
The larger scale data (shown in Figures \ref{fig:overlay_B}a and \ref{fig:overlay_B}b) are from the single dish by CSO \citep[Hertz,][]{Dowell+etal_1998} at 350 $\mu$m \citep[][]{Dotson+etal_2010} at a scale of 3 pc
and by JCMT \citep[SCUBA,][]{Holland+etal_1998} at 850 $\mu$m \citep{Chrysostomou+etal_2002} at a scale of 2 pc.
The smaller scale data are from our SMA-SubC at a scale of 0.3 pc and from SMA-Ext at a scale of 60 mpc at 870 $\mu$m (Figure \ref{fig:overlay_B}c). 
The position angle of the magnetic field ($\phi_{\rm B}$) is obtained by rotating the detected $\phi_{\rm pol}$ by 90$\degr$. 
We also provide a quantitative analysis in support of the observational
description in section \ref{sec:field_analysis}.  We conclude in section 
\ref{sec:global_picture} with a schematic scenario of the magnetic field observations
so far in the W51 North region. 

%
\subsection{SMA high-resolution magnetic field morphologies}
                                          \label{sec:field_structure}

The dust continuum morphology in the SMA-SubC 
shows regular contours with an extension to the west. These contours show 
some symmetries along an EW axis around
$\Delta \rm Dec.$=0$\arcsec$ (Figure \ref{fig:overlay_B}c). Similarly, the detected polarization
segments reveal a north-south symmetry (upper and lower planes) along the 
same axis. Assuming flux-freezing with the gas mostly moving along the field 
lines, the combined observed symmetry patterns (in the dust continuum and 
the field morphologies) are suggestive to find mass accumulations mostly
along the symmetry axis. Indeed, the detected cores (SMA1 to SMA4) in the SMA-Ext data
(Figure \ref{fig:overlay_B}c) are mostly aligned on an EW axis at the above declination. Thus, 
the observed field configuration on the $\sim 0.3$~pc scale (SMA-SubC) 
appears to be channeling and aligning material, which is detected as denser
cores on a $\sim 60$~mpc scale (SMA-Ext) once the extended emission is filtered out.

Unlike the symmetric field configuration on the $\sim 0.3$~pc scale, the smaller
individual cores on the $\sim 60$~mpc scale show very different field morphologies. 
SMA1 exhibits $\phi_{\rm B}$ orientations roughly comparable to that of the one corresponding segment of the 
subcompact data (Figure \ref{fig:overlay_B}c). 
The morphology possibly also shows a hint of gravity dragging in the B field lines toward the center.
The patch of segments in SMA2
is rather orthogonal to the SMA-SubC segment next to it. 
SMA3 and SMA4 show again a more closer alignment with the 
single SMA-SubC segment but with some deviations up to about 45$^{\circ}$.
Thus, within one resolution element of the SMA-SubC data (where neighboring
segments along the symmetry axis show only little changes in orientation), the 
detected $\phi_{\rm B}$ in the SMA-Ext data show changes in 
orientation of up to 90$^{\circ}$. 

The B morphology associated with SMA1 is particularly interesting.
As mentioned in section \ref{sec:introduction}, 
collapsing signatures and a rotating molecular ring have been reported toward SMA1 \citep{Sollins+etal_2004,Zapata+etal_2008,Zapata+etal_2009}.
The detected pinched B field morphology associated with SMA1 is, therefore, consistent with the picture of the B field lines being dragged by inward motions and/or the rotation of the molecular ring. 

Despite the very different field orientations in SMA1 to SMA4, the $\phi_{\rm B}$ in these cores all reveal a common trend: the field segments tend to be aligned with 
the orientations of the dust continuum intensity gradients, $\phi_{\rm \nabla I}$, in many locations\footnote{
The dust intensity gradients are orthogonal to the dust emission contours (in the plane of sky).
The resulting position angles, $\phi_{\rm \nabla I}$, are defined in the range of 0$\degr$ to 180$\degr$.
}. 
In section \ref{sec:field_analysis}, we will have a more quantitative analysis on this correlation.

%
\subsection{Quantitative Analysis}\label{sec:field_analysis}

Figures \ref{fig:analysis}a to \ref{fig:analysis}d show the dust emission maps overlaid with the magnetic field segments (red) and the intensity gradient segments (blue).
Inspired by the close alignment of the position angles of the intensity gradients, $\phi_{\rm \nabla I}$, and the position angles of the B field, $\phi_{\rm B}$, we apply further analyses in this subsection.

%
\begin{deluxetable*}{l|ccc|ccc|c|ccc}{!th}  
\tabletypesize{\scriptsize}
\tablewidth{0pt}
\tablecaption{Analysis Summary
                      \label{tab:analysis}}

\tablehead{
\colhead{Parameters}  & \multicolumn{3}{c}{CSO}  & \multicolumn{3}{c}{JCMT} & \colhead{SMA-SubC} & \multicolumn{3}{c}{SMA-Ext}
}

\startdata

$\theta_{\rm (syn)}$ ($\arcsec$) & \multicolumn{3}{|c|}{20} & \multicolumn{3}{|c|}{9.3} & 4 & \multicolumn{3}{|c}{0.7} \\

$\lambda$ ($\mu$m) & \multicolumn{3}{|c|}{350} & \multicolumn{3}{|c|}{850} & 870 & \multicolumn{3}{|c}{870} \\ 

B morphology & \multicolumn{3}{|c|}{uniform} & \multicolumn{3}{|p{3.4cm}|}{uniform patches, changes in large scale} & \multicolumn{1}{|p{3cm}|}{shaped, cometary-like, N-S mirror symmetry} & \multicolumn{3}{|c}{shaped, possibly hourglass-like} \\ \hline

 & all & core$^{a}$ & out$^{b}$ & all & core$^{a}$ & out$^{b}$ & all & all & core1$^{c}$ & core2$^{c}$ \\ \hline

$\mathcal{C}$           & 0.81 & 0.88 & 0.59  & 0.70 & 0.73 & 0.66 & 0.78 & 0.88 & 0.89 & 0.86 \\ 
$\sigma_{\mathcal{C}}$     & 0.06 & 0.08 & 0.08  & 0.03 & 0.04 & 0.03 & 0.05 & 0.02 & 0.02 & 0.03 \\
$\langle|\Delta \rm \phi_{\rm B}|\rangle$  & 12\degr & - & - & 43\degr & - & - & 43\degr & 39\degr & - & - \\ 

$\sigma_{|\Delta \rm \phi_{\rm B}|}$  & 13\degr & - & - & 27\degr & - & - & 26\degr & 26\degr &  - & - \\ 

$\langle|\delta|\rangle$  & 40\degr & 51\degr & 34\degr & 44\degr & 38\degr & 48\degr & 37\degr & 35\degr & 34\degr & 36\degr\\ 

$\sigma_{|\delta|}$      & 22\degr & 29\degr & 15\degr & 27\degr & 28\degr & 26\degr & 26\degr & 24\degr & 26\degr & 23\degr \\ 

$\langle\Sigma_{\rm B}\rangle$ & 0.71 & 0.69(0.59) & 0.72 & 1.17 & 0.42(0.27) & 1.84 & 
0.47(0.18) & 0.48(0.38) & 0.47(0.37) & 0.48(0.39)

\enddata

\tablecomments{Statistical quantities derived using the position angles of the intensity gradient ($\phi_{\rm \nabla I}$) and the position angles of the magnetic field ($\phi_{\rm B}$). 
$\theta_{\rm (syn)}$ and $\lambda$ are the angular resolution and the wavelength of the traced emission, respectively. 
$\mathcal{C}$ is the Pearson's correlation coefficient between $\phi_{\rm B}$ and $\phi_{\rm \nabla I}$, and its standard deviation is given by $\sigma_{\mathcal{C}}$.
$\langle|\triangle \phi_{\rm B}|\rangle$ and $\sigma_{|\triangle \phi_{\rm B}|}$ are the mean difference of $\phi_{\rm B}$ of polarization pairs and the standard deviation of the difference, respectively.
$\langle|\delta|\rangle$ and $\sigma_{|\delta|}$ is the mean of the absolute difference between $\phi_{\rm B}$ and $\phi_{\rm \nabla I}$ and its standard deviation, respectively.
$\langle\Sigma_{\rm B}\rangle$ is the mean of the ratio of the magnetic field force to gravitational force ($\Sigma_{\rm B}$), which is calculated with the method introduced in \citet{Koch+etal_2012a}. Values after removing the outliers are listed in parenthesis.
}
\tablenotetext{(a)}{Values derived using the pixels marked within the cyan square in Figure \ref{fig:analysis}.}

\tablenotetext{(b)}{Values derived using the pixels outside the cyan square in Figure \ref{fig:analysis} (named as outside of the core region in the text).}

\tablenotetext{(c)}{Core1 and core2 refers to SMA1,2 and SMA3,4, respectively.}

\end{deluxetable*}
%
%

%
\subsubsection{Correlation of $\phi_{\nabla \rm I}$ and $\phi_{\rm B}$}\label{sec:correlation}
\begin{figure*}
\begin{center}
\includegraphics[scale=1]{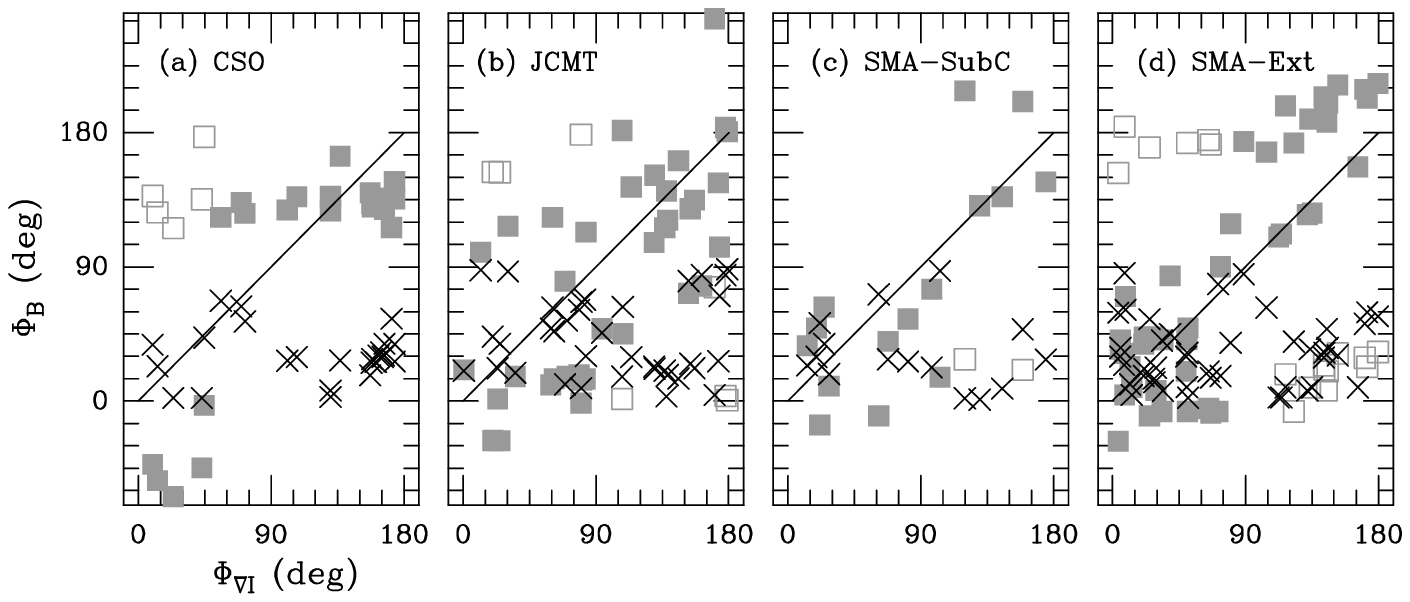}
\caption{Position angles of the intensity gradient orientations, $\phi_{\nabla \rm I}$, and the B field orientations, $\phi_{\rm B}$, in 
solid squares.  Crosses mark the absolute differences between $\phi_{\nabla \rm I}$ and $\phi_{\rm B}$. 
In order to properly display their correlations, the $\phi_{\rm B}$ is re-defined beyond the 0$\degr$-180$\degr$ range
(filled squares above 180$\degr$ and below 0$\degr$) for ambiguous situations; i.e., a pair of position angles of 
175$\degr$ and 5$\degr$ is displayed as  175$\degr$ and 185$\degr$. 
The hallowed squares mark their original data points. 
Note that this re-definition does not change the correlation in any way, but is merely a consequence of the 180$\degr$
ambiguity in the range of position angles. 
The straight line represents a perfect correlation.}
\label{fig:plot_gradi_b}
\end{center}
\end{figure*}

We first compare the correlation of $\phi_{\nabla \rm I}$ and $\phi_{\rm B}$ in these 4 data sets.
The correlation plots are shown in Figure \ref{fig:plot_gradi_b}.
The uniformity of the B field traced with the CSO is apparent here with a large group of data points scattering around $\phi_{\rm B}$ of 120$\degr$.
Except for the CSO data, the correlation between $\phi_{\nabla \rm I}$ and $\phi_{\rm B}$ can be seen by eye.

%
%

The correlation is further analyzed quantitatively using Pearson's correlation coefficient, $\mathcal{C}$. 
The uncertainties in the correlation coefficient (the standard deviation), $\sigma_{\mathcal{C}}$, are estimated using a Monte-Carlo approach where errors in both $\phi_{\rm B}$ and $\phi_{\rm \nabla I}$ are propagated through the calculation. Gaussian distributions 
of the $\phi_{\rm B}$ (= $\phi_{\rm pol}+90\degr$) uncertainties at two sigma level, i.e.  2 times the $\sigma_{\rm \phi_{\rm pol}}$ term in Table \ref{tab:pol_data}, are adopted. Gaussian distributions with mean uncertainties of $\sim 3\degr - 5^{\circ}$ --
which is the result after interpolating between Stokes $I$ values in order to calculate gradients --
are used for the $\phi_{\nabla \rm I}$ uncertainties. 
Resulting values of $\sigma_{\mathcal{C}}$ are in the range of 0.02 to 0.08.

As listed in Table \ref{tab:analysis}, $\mathcal{C}$ is highest in the highest resolution data (SMA-Ext), being 0.88$\pm$0.01, suggesting that the correlation between  $\phi_{\nabla \rm I}$ and $\phi_{\rm B}$ is tightest in the structures revealed with SMA-Ext.
For SMA-SubC, the $\mathcal{C}$ value is 0.78$\pm$0.03.
The $\mathcal{C}$ values for the CSO and the JCMT data are 0.81$\pm 0.03$ and 0.70$\pm 0.02$, respectively. 
We further derive the $\mathcal{C}$ values for the CSO and the JCMT data for two separated regions: the core and the region outside of the core. 
The $\mathcal{C}$ values are larger in the cores, being 0.88$\pm$0.04 and 0.73$\pm$0.02 in the CSO and JCMT data, respectively. This is similar to the high $\mathcal{C}$ value in SMA-Ext, further confirming that the ${\phi_{\nabla \rm I}}$ and $\phi_{\rm B}$ correlation is higher in the cores.
Outside of the core region, the $\mathcal{C}$ values are generally smaller (0.59$\pm$0.04 for CSO and 0.66$\pm$0.02 for JCMT), and the correlation is worse. 
It is systematic that both the CSO and the JCMT data show a difference in $\mathcal{C}$ between the core and the outside-the-core region. 
We will propose a physical interpretation for such a change in $\mathcal{C}$ in section \ref{sec:pi_method}.

\begin{figure*}[!ht]
\begin{center}
\includegraphics[scale=1]{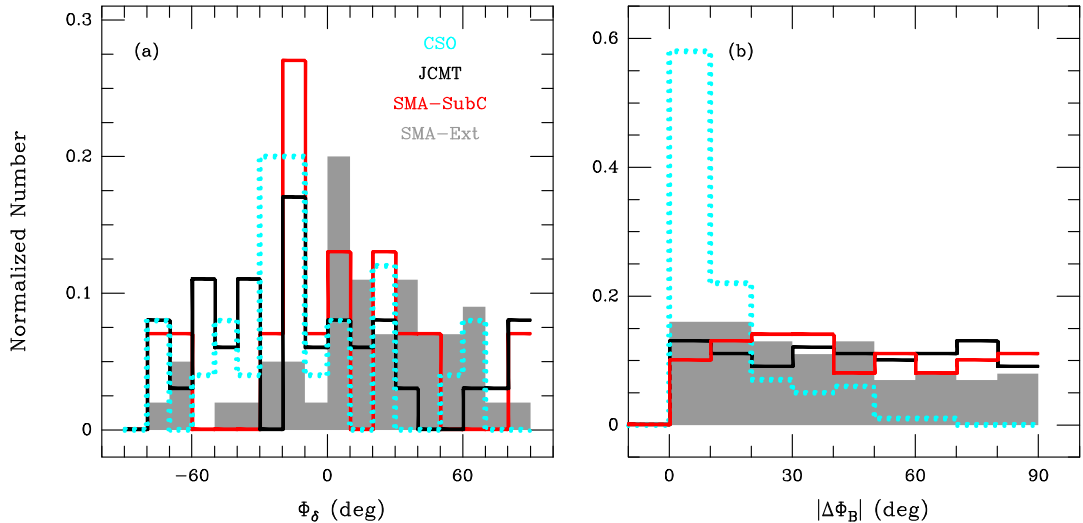}
\caption{(a) Histograms of the differences in position angles between the B field and the intensity gradient ($\phi_{\rm \delta}$). 
(b): Histograms of the absolute values of the differences in position angles of the polarization pairs ($|\triangle \phi_{\rm B}|$).}
\label{fig:corr_gradI_B}
\end{center}
\end{figure*}

In addition to the $\phi_{\nabla \rm I}$-$\phi_{\rm B}$ plot, we also compare the distribution of the angle difference, $\phi_\delta$, between $\phi_{\rm B}$ and $\phi_{\nabla \rm I}$ (Figure \ref{fig:corr_gradI_B}a).
It is important to note that all the distributions are 
non-Gaussian, i.e., the differences in the magnetic field and intensity gradient
orientations are not only due to a measurement uncertainty, but point toward
a more fundamental physical interpretation as explored in \citet{Koch+etal_2012a}.

The differences of individual pairs of field segments over an observed map are also compared. 
The histograms in Figure \ref{fig:corr_gradI_B}b display
$|\Delta\phi_{\rm B}|\equiv |\phi_{\rm B,i}-\phi_{\rm B,j}|$ for all possible pairs of segments $i$ and $j$ in a map. 
A systematic difference is apparent between the CSO data and the other 3 data sets. 
The CSO distribution peaks at 
small values, with an average change in orientation $\langle|\Delta\phi_{\rm B}|\rangle=12^{\circ}$,
whereas the other distributions are close to flat with average changes around
$40^{\circ}$ (Table \ref{tab:analysis}). This reflects the visual 
impression that the field is mostly uniform in orientation in the CSO map
at $\phi_{\rm B}$ of 120$\degr$. In the other maps, the field orientations 
seem to be more significantly shaped and overcome by other forces, erasing the 
original likely uniform orientation of the field. Thus, one might conclude that 
on the largest scales, a uniform field morphology points to a more significant 
role of the magnetic field. 

%
%

%
\subsubsection{Application of the Polarization - Intensity Gradient Method}\label{sec:pi_method}

In this subsection, we apply the new polarization-intensity gradient method to derive the force ratio ($\Sigma_{\rm B}$) map of the magnetic field to the gravitational 
force \citep[see][for the development of the method]{Koch+etal_2012a}.
The derivation of $\Sigma_{\rm B}$ makes use of $|\phi_{\rm \delta}|$ in combination with an additional angle between $\phi_{\rm \nabla I}$ and the local gravity direction.
This local gravity direction is calculated from the positions of the gravity centers which are identified with the emission peaks in a map\footnote{
For the original derivation of the force ratio $\Sigma_{\rm B}$ we refer the reader to \citet{Koch+etal_2012a}.
$\Sigma_{\rm B}$ is found from an MHD force equation, comparing the field tension force $F_{\rm B}$ and the gravitational force $F_{\rm G}$ via 
measurable angles: $\Sigma_{\rm B}=F_{\rm B}/F_{\rm G}=\sin\psi / \sin\alpha$. The angle $\psi$ is defined by the intensity gradient and the direction of gravitational force. The angle $\alpha$ is between the intensity gradient and the originally detected polarization orientations.}.
For SMA-Ext, two gravity centers are defined, namely SMA1 and SMA3, around which most of the polarization is detected.
Figures \ref{fig:analysis}e to \ref{fig:analysis}h show 
maps of the force ratio $\Sigma_B$. 
Values averaged over the map, 
$\langle\Sigma_{\rm B}\rangle$, are listed in Table \ref{tab:analysis}. 
The uncertainties of the derived ratios depend on several factors, such as the $\phi_{\rm pol}$ uncertainties and the positions of the gravity centers in a map.
Adopting uncertainties like in the previous section for $\mathcal{C}_{\rm err}$, typical errors in $\Sigma_{\rm B}$ are $\sim 5\%$ for core-like structures like in SMA-Ext. The errors can grow to $\sim 20\%$ for larger scale field morphologies where gravity centers are less defined. 
Averaging $\Sigma_{\rm B}$ over a larger area (cores, outside-core regions) will typically lead to errors of a few percents.

The $\langle\Sigma_{\rm B}\rangle$ values of the large-scale observations with the CSO and the JCMT are around 0.71 and 1.17 
when averaged over the entire area, respectively. 
If we separate the values into core region and outside-the-core region, $\langle\Sigma_{\rm B}\rangle$ is 0.69 ($\sim$ 0.59 if outliers are removed) and 0.72 for CSO in the core and outside, respectively. 
For JCMT, the variation is more dramatic, being 0.42 (0.27 if outliers are removed) and 1.84 for the core and the region outside of the core, respectively. We, thus, find that 
$\Sigma_{\rm B}$ tends to be smaller for the cores than in the areas outside of the cores.
The smaller-scale SMA data also show a reduced magnetic field significance with $\langle\Sigma_{\rm B}\rangle\sim 0.5$. 
Generally, closer to and around the emission peaks, the force ratios seem to be smaller.
This implies that on these scales the magnetic field is more shaped and dominated by gravity.
In the regions outside of the core, the ratios are larger or the field
is even dominating gravity ($\Sigma_{\rm B} > 1$ for JCMT).

This change in the role of the magnetic field based on the force ratio $\Sigma_{\rm B}$ might provide a physical explanation of the change of the correlation coefficient $\mathcal{C}$ described in section \ref{sec:correlation}. 
The tighter correlation of $\phi_{\nabla \rm I}$ and $\phi_{\rm B}$ in the cores suggests that the B field lines are mainly dragged by gravity and, consequently, align with the intensity gradient.
Therefore, quantitatively, the B field is indeed of minor importance in the cores.
We emphasize that the $\mathcal{C}$ value is purely statistical without any physical assumptions. 
In contrast, the $\Sigma_{\rm B}$ value is calculated by identifying various terms in an ideal MHD force equation in an observed Stokes $I$ and dust polarization map.
These two approaches are, thus, largely independent.
They, nevertheless, seem to lead to a consistent picture
where the magnetic field significance is reduced in the core regions.

%
\begin{figure*} [!ht]
\begin{center}
\includegraphics[scale=0.4]{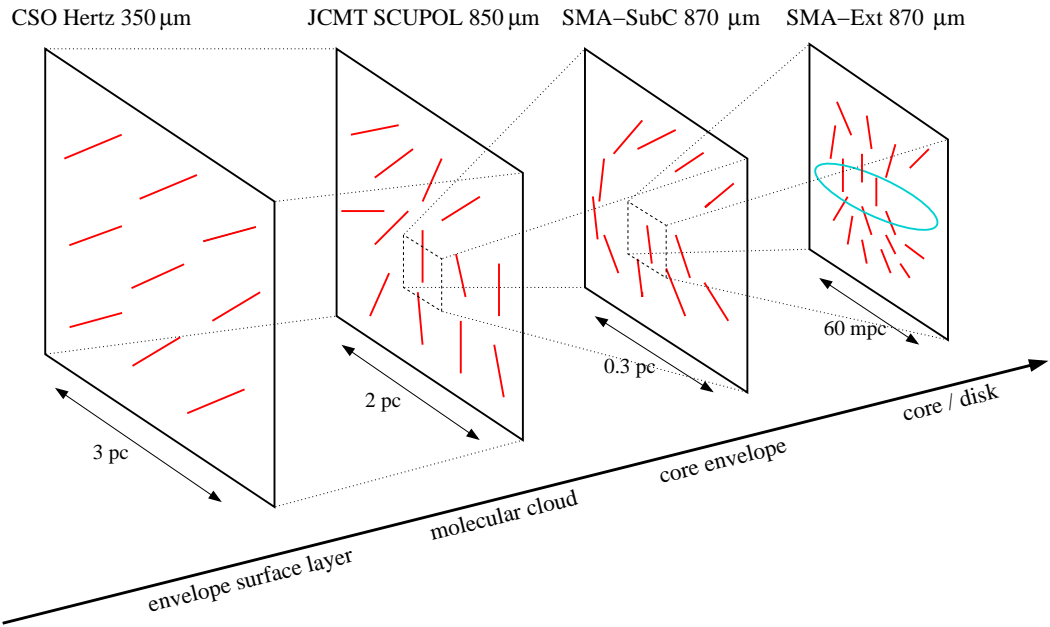}
\caption{\label{fig:schematic}
Schematic illustration of the different magnetic field structures
sampled at various depths along the line of sight and at different scales.
The cyan ellipse marks the location of the collapsing and rotating molecular core found in \citet{Zapata+etal_2009}.
}
\end{center}
\end{figure*}
%

%
\subsection{Global Picture of the Magnetic Field in W51 North}  \label{sec:global_picture}

Figure \ref{fig:schematic} schematically summarizes the 4 magnetic field observations analyzed in the 
previous sections. Each panel represents a plane-of-sky projected observation of the CSO, JCMT, SMA-SubC
and SMA-Ext, respectively, where a subsequently higher-angular-resolution observation is zooming in on a previous one.
In this sequence of observations, field morphologies are sampled from a 3 pc scale down to a 60 mpc physical scale.

These four panels represent the field geometry from different layers or from different scales in the W51 North region.
For the CSO image, the flux from an optically thick 20$\arcsec$ (the CSO resolution) source at a rotational temperature T of 15 K will be 270 Jy, and 760 Jy at T=30 K. 
This is in the range of Stokes I intensity in the observed region \citep{Dotson+etal_2010}. 
We note that the rotational temperature of W51 N is 45 K as traced by ammonia lines \citep{Zhang_Ho1997} and the detected structure is $\sim$10$\arcsec$. A decreasing temperature from the UCHII region to the outer surrounding gas is further proposed by \citet{Zhang+etal_1998}. 
Thus, the gas temperature within a 20$\arcsec$ beam is unlikely to be much higher than 30 K.
Unless the temperatures are substantially higher, the average opacity must be of order 1.  
Furthermore, the observed polarization arises from the difference in emissivity of dust grains. 
In an optically thick medium, the polarization comes from upper layers.
We then expect the field morphology to rather result from an upper surface layer than from the deeper embedded core.
This is the first panel in Figure \ref{fig:schematic} which is labeled as {\it envelope surface layer}.
The field is mostly uniform here, with a single orientation still reflecting the largest scale field structure. 
The field-to-gravity force ratio is relatively large ($\sim 0.7$, Table \ref{tab:analysis}) and rather uniform 
with only a minor difference 
between core and outside-core region (Figure \ref{fig:analysis}e).  The correlation $\mathcal{C}$, between field and 
intensity gradient orientations, is poorest outside the core ($\mathcal{C}\sim 0.59$). This all is suggestive of a magnetic 
field that is still retaining its tension force in this surface layer.  

The JCMT 850 $\mu$m observation -- with a 
higher resolution and reaching deeper inside the core -- shows still some uniform patches in field morphology, 
but also clearly displays changing field orientations in the core. In this next deeper layer (labeled as 
{\it molecular cloud} in Figure \ref{fig:schematic}), the field ratio shows a clear difference between core ($\sim 0.4$)
and outside-core region ($>1$, Figure \ref{fig:analysis}f). This might indicate that gravity is overcoming the 
field tension in the core area, whereas the field remains significant at larger distances, possibly even holding 
back material from further accreting onto the core. Thus, the CSO and JCMT observations, despite covering 
similar areas, seem to sample different depths where the magnetic field seems to play different roles. 

The SMA-SubC 870 $\mu$m observation is akin to zooming in on the JCMT observation. The force ratio $\Sigma_B$ 
and the correlation $\mathcal{C}$ are comparable to the JCMT values. This observation seems to further sharpen the JCMT 
core, removing some of the extended emission (third panel labeled as {\it core envelope} in 
Figure \ref{fig:schematic}). In addition to the JCMT observation (which already shows a core being dominated
by gravity), the North-South symmetry in the SMA-SubC field morphology provides a hint for yet another role
of the magnetic field: material can be channeled along the field lines onto an east-west axis. 
Finally, the SMA-Ext data with sub-arcsecond resolution clearly resolve the $\sim 0.3$pc  structure into 
4 cores (fourth panel labeled as {\it core/disk} in Figure \ref{fig:schematic}). Distinct morphologies are
found for individual cores, with all cores consistently showing the tightest correlations ($\sim 0.9$) and 
clearly low force ratios ($\sim 0.4$). At this scale, an hourglass-like morphology is becoming visible. 

In summary, we thus propose a schematic scenario in Figure \ref{fig:schematic}, for the change in the 
B field, from the largest scale envelope surface layer down to the scale where a collapsing core is being revealed.
The scenario described above emerges from visual impressions from the field morphologies, combined 
with a quantitative analysis from correlations and force ratios. 

Finally, it is interesting to remark that {\it generally} 
the correlation coefficients $\mathcal{C}$ are found to be larger in the cores than outside of the cores in all 4
scales and/or layers.
The mean $\phi_{\delta}$ (the difference between $\phi_{\rm \nabla I}$ and $\phi_{\rm B}$) and its standard deviation in the cores tend to be smaller than the ones outside the cores. 
Both the analyses of $\mathcal{C}$ and $\phi_{\delta}$ suggest that $\phi_{\rm B}$ has a tighter correlation with $\phi_{\rm \nabla I}$ in the cores. 
Similarly, the force ratio of the B field strength and the gravity ($\Sigma_{\rm B}$) in these 4 data sets also reveal a common trend: 
$\Sigma_{\rm B}$ tends to be smaller in the cores than outside the cores, suggesting that gravity is 
generally playing a more important role within the cores. 
As already quantified and observed for other sources \citep{koch12b}, we can thus consistently confirm that the magnetic 
field significance systematically decreases toward the cores for various physical scales.

%

%
%
\section{Conclusion and Summary}\label{sec:summary}
 
We present new high-resolution SMA dust polarization observations toward the molecular cloud W51 North.
Beyond the polarization imaging we also apply new analysis techniques providing insight into the role
and importance of the magnetic field. 
Furthermore, we make the first attempt to present 
a magnetic field picture from the largest scales (3~pc) -- from previously published CSO and JCMT data --
to the SMA detected collapsing cores ($\sim$ 60~mpc).
The key points are summarized in the following.
 
\begin{enumerate}

\item Our new SMA high-resolution observations in the subcompact (SMA-SubC, angular resolution $\theta_{\rm syn}$ of 4$\arcsec$) and extended (SMA-Ext, $\theta_{\rm syn}$ of $0\farcs$7)
array  configurations clearly detect and resolve the 
dense structures in W51 North. With the observed thermal dust Stokes $I$ continuum emission at a wavelength of 870 $\mu$m, four dense cores, SMA1 to SMA4, are resolved in an east-west dust ridge.

\item The polarized thermal continuum at 870 $\mu$m is detected and resolved with an angular resolution 
up to 0$\farcs$7 for the first time with the SMA. 
The polarization percentage $P$ ranges from 1 \% to 4 \%. 
$P$ is found to decrease with higher intensity for both the SMA-SubC and SMA-Ext data.
This trend is also largely identical to the larger scale CSO and JCMT polarization data.

\item The observed magnetic field configuration on the $\sim 0.3$~pc scale with the SMA-SubC shows some 
mirror-symmetry features from North to South. Here, the field lines might be channeling and aligning material from 
both North and South. At a $\sim 60$~mpc scale resolution observed with the SMA-Ext, dense cores are, indeed, detected along the east-west 
mirror symmetry axis.
Smooth changes in field morphologies are found within individual cores, whereas large overall changes in position angles 
can occur in between the cores.

\item 
A correlation between dust intensity gradient position angles 
($\phi_{\rm \nabla I}$) and magnetic field position angles ($\phi_{\rm B}$) is found in the CSO, JCMT and both SMA data sets. 
This correlation is further analyzed quantitatively. 
A systematically tighter correlation between $\phi_{\rm \nabla I}$ and $\phi_{\rm B}$ is found in the cores, whereas
at larger distances the correlation decreases. 

\item Magnetic field to gravity force ratio ($\Sigma_{\rm B}$) maps are derived using the newly developed polarization - intensity
gradient method \citep{Koch+etal_2012a}. 
In this method, $\Sigma_{\rm B}$ is quantified by measuring angles that reflect the resulting forces shaping 
the dust Stokes $I$ and magnetic field morphologies.
We find that the force ratios tend to be small ($\Sigma_{\rm B}\simlt 0.5$) in the cores in all 4 data sets. 
In regions outside of the cores, the ratios increase or the field
is even dominating gravity ($\Sigma_{\rm B} > 1$).
This possibly provides a physical explanation of the tightening correlation between
$\phi_{\nabla \rm I}$ and $\phi_{\rm B}$ in the cores: the more the B field lines are dragged and aligned by gravity, 
the tighter the correlation is. Intuitively, in such a scenario one expects gravity to play an increasingly dominating
role. The analysis via the force ratio $\Sigma_{\rm B}$ seems to consistently confirm this expectation. 
We stress that correlation and $\Sigma_{\rm B}$ are independent approaches. They, nevertheless, seem to reveal a 
matching picture.
  

\item 
Interpreting the four polarization observations covering different physical scales, we propose a schematic 
scenario for the magnetic field in W51 North (Figure \ref{fig:schematic}).  In the largest envelope surface layer
(CSO), the field has kept much of its tension force. In the deeper-layer JCMT observation, gravity has overcome
the magnetic  field in the core whereas in the outside the field tension is significant, capable of holding 
back material against gravity. Both higher-resolution SMA observations show gravity dominating the field 
tension. In SMA-SubC, the field is possibly channeling material from North and South onto an east-west axis.
In SMA-Ext, four higher-resolution cores are aligned in an east-west axis, with a field morphology  pointing toward hourglass-like structures as expected in a core/disk system.

\end{enumerate}

This research was supported by NSC grants NSC99-2119-M-001-002-MY4 and NSC98-2119-M-001-024-MY4. This work was supported by ÒProgramme National de Physique StellaireÓ (PNPS) and ÒProgramme National de Physique Chimie du Milieu InterstellaireÓ (PCMI) from INSU/CNRS.

\bibliographystyle{apj}                       
\bibliography{w51n}

\end{document}